\documentclass{jkpaper}

\usepackage{tikz}
\usetikzlibrary{decorations.markings}
\usetikzlibrary{decorations.pathreplacing}
\usetikzlibrary{arrows.meta}

\title{Subregions, Minimal Surfaces, and Entropy in Semiclassical Gravity}
\author{Josh Kirklin}
\email{jjvk2@cam.ac.uk}
\institution{Department of Applied Mathematics and Theoretical Physics, Centre for Mathematical Sciences, University of Cambridge, Cambridge, UK}

\abstr{%
		For a large class of density matrices in semiclassical gravity, it is shown that the reduced density matrix which corresponds to tracing over the degrees of freedom in a spatial subregion is dominated by states for which the area of the boundary of the subregion is minimised. In the semiclassical limit, the entropy of the reduced density matrix is found to have a leading order contribution equal to one quarter of the minimal area in natural units. This is consistent with the Ryu-Takayanagi conjecture. An extension to higher derivative theories of gravity is established, for which the area is replaced by a dynamical generalisation of the Wald entropy. %
}

\bibliography{refs}

\newcommand{\citegeneralisedRT}{\cite{Fursaev:2013fta,Bhattacharyya:2013jma,Dong:2013qoa,Camps:2013zua,Camps:2014voa,Bhattacharyya:2014yga,Miao:2014nxa}}
\newcommand{\citecovariant}{\cite{REGGE1974286,Crnkovic,300yearscrnkovicwitten,Lee:1990nz,Wald:1993nt,Iyer:1994ys,Barnich:2001jy}}
\newcommand{\citeambiguities}{\cite{Jacobson:1993vj,Iyer:1994ys}}

\begin{document}

\maketitleandtoc

\section{Introduction}
\label{Section: Introduction}

The first serious insight into the holographic nature of quantum gravity was arguably the realisation by Hawking and Bekenstein that the entropy of a black hole in general relativity is proportional to the area of a cross-section of its event horizon~\cite{Hawking:1971tu,Bekenstein:1973ur,Hawking:1976de}. Thermodynamical entropy is usually an extensive quantity, meaning it scales with the volume of space. The fact that the black hole entropy scales like an area, which has dimensionality one lower than a volume, is strongly suggestive that the microscopic degrees of freedom present in quantum gravity can be understood as living on a manifold with dimension one less than that of the bulk spacetime~\cite{tHooft:1993dmi,Bousso:2002ju}.

In~\cite{Wald:1993nt}, Wald generalised the first law of black hole mechanics~\cite{Bardeen:1973gs} to arbitrary theories of gravity. This led him to suggest a value for the black hole entropy in such theories. Let $Q$ be the Noether charge associated with diffeomorphisms along the Killing vector field generating the black hole horizon. He showed that $Q$ has two contributions: one at infinity, and the other at the bifurcation surface $\mathcal{S}$. The black hole entropy is then a certain multiple of the contribution at $\mathcal{S}$. In terms of the Lagrangian density $L$, a simple formula for it is
\begin{equation}
	S_{\text{Wald}} = -\frac{2\pi}{\hbar}\int_{\mathcal{S}}\dd[D-2]{\sigma}\sqrt{\det q}\epsilon_{ab}\epsilon_{cd}\fdv{L}{R_{abcd}},
	\label{Equation: Wald entropy}
\end{equation}
where $\sigma$ are a set of coordinates on $\mathcal{S}$, $q_{ab}$ is the induced metric on $\mathcal{S}$, $\epsilon_{ab}$ is the binormal to $\mathcal{S}$, and $\fdv{L}{R_{abcd}}$ is the Euler-Lagrange derivative of $L$ with respect to the Riemann tensor $R_{abcd}$. This quantity is usually called the Wald entropy.

The derivation of the Wald entropy only works for stationary black holes. In~\cite{Iyer:1994ys}, Iyer and Wald proposed a generalisation to the dynamical case. Let $\phi$ represent the dynamical fields in spacetime. Iyer and Wald gave a definition for the `boost-invariant' part $\bar\phi$ of the fields $\phi$, and proposed that the appropriate entropy to associate with a slice of a dynamical black hole horizon should be found by evaluating the Wald entropy over that slice, after having made the replacement $\phi\to\bar\phi$. They came to this conclusion by requiring that the entropy satisfy a set of conditions motivated by the thermodynamical interpretation. The resulting definition for the entropy of a dynamical black hole horizon is usually called the Iyer-Wald dynamical entropy. We will denote it $S_{\text{Iyer-Wald}}$.

A large number of other relations between the areas of various surfaces, and the entropies of certain states associated with them, have been proposed. Some of these proposals are just conjectures, while others are mathematically proven statements, and they form a part of the currently ongoing exploration of the role of quantum information in gravity. While this is a subject that is still not fully understood, it is clear that one can learn much about the structure of quantum gravity from information-theoretic considerations. A helpful introduction to many of these ideas can be found in~\cite{Harlow:2014yka,VanRaamsdonk:2016exw,Harlow:2018fse}.

One particular area-entropy relation in the context of the AdS/CFT duality~\cite{Maldacena:1997re,Witten:1998qj} is known as the Ryu-Takayanagi conjecture~\cite{Ryu:2006bv,Ryu:2006ef,Hubeny:2007xt,Rangamani:2016dms}. One considers a density matrix $\rho$ on the CFT side, which is defined over a spatial slice $B$ of the conformal boundary of the bulk AdS spacetime. A subregion $T\subset B$ has complement $\bar{T}\subset B$, and we can define the reduced density matrix $\rho_T$ by taking the partial trace of $\rho$ over all of the degrees of freedom present in $\bar{T}$. The von Neumann entropy of the region $T$ is then given by $S_T = -\tr(\hat\rho_T\log\hat\rho_T)$, where $\hat\rho_T$ is the normalised reduced density matrix. The Ryu-Takayanagi conjecture claims that this entropy should contain a contribution equal to $\frac{\mathcal{A}_{\text{min}}}{4G\hbar}$, where $\mathcal{A}_{\text{min}}$ is the smallest possible area of a codimension 2 surface in the bulk which shares its boundary with $T$. This codimension 2 surface is known as the `entangling' surface. 

The evidence for the Ryu-Takayanagi conjecture is plentiful~\cite{Calabrese:2004eu,Hartman:2013mia,Faulkner:2013yia,Casini:2011kv,Faulkner:2014jva}. There are also more general versions of the conjecture that are supposed to apply to classes of higher derivative theories of gravity~\citegeneralisedRT, in which the area is replaced by some other functional on the codimension 2 bulk surface. One particularly enticing piece of evidence for these generalised entropies is the claim in~\cite{Wall:2015raa} that they are to a certain extent the only entropies which obey the second law at the linearised level.

Maldacena and Lewkowycz provided a heuristic proof~\cite{Lewkowycz:2013nqa} of the Ryu-Takayanagi conjecture that applies in certain cases. Their argument is based on what is known as the `replica trick'. For each integer $n$, one can calculate a Renyi entropy $S_T[n]=\frac1{1-n}\log\tr(\hat\rho^n_T)$ of the reduced density matrix, and it turns out that this calculation is sometimes simpler than the von Neumann one. In the Maldacena-Lewkowycz case, this entails a calculation of the semi-classical partition function in the presence of a conical defect. Under the assumption that the Renyi entropy can be continued consistently to non-integer $n$ in a neighbourhood of $n=1$, one can then calculate the von Neumann entropy with the formula
\begin{equation}
	-\big[(1-n)\partial_nS_T[n]+S_T[n]\big]_{n=1} = -\left[\frac1{\tr(\hat\rho_T^n)}\tr(\hat\rho_T^n\log\hat\rho_T)\right]_{n=1} = -\tr(\hat\rho_T\log\hat\rho_T)=S_T.
\end{equation}
Maldacena and Lewkowycz showed that when one calculates $S_T$ in this way, an equation of motion for the entangling surface naturally arises, and this equation of motion exactly picks out surfaces which extremise its area. Moreover, they found the resulting entropy to be given by the Ryu-Takayanagi formula.

In this paper we will show that the Ryu-Takayanagi conjecture holds for a large class of density matrices in the semiclassical limit. We will use a method which, unlike that of Maldacena and Lewkowycz, does not require use of the replica trick, including the assumption that the continuation to non-integer $n$ is consistent, and does not necessitate the introduction of conical defects. We will work in the gravity side of the holographic duality, and in fact our arguments make no reference to the duality whatsoever. In the course of the paper we will show that the reduced density matrix itself is dominated by elements for which the entangling surface has minimal area, which is a somewhat stronger result than the original claim of Ryu and Takayanagi. The argument works for any choice of boundary conditions, and so in particular does not just apply to asymptotically AdS spacetimes. In addition, the extension to more general theories is relatively clear. The result is the same, except that the area is replaced by a dynamical version of the Wald entropy (not necessarily the same as the Iyer-Wald version).

In gravity, there is an apparent gauge ambiguity in the definition of subregions, arising from the expectation that diffeomorphism invariance in the bulk allows one to arbitrarily deform the entangling surface on the boundary of the subregion, and not incur any physical consequences. An attempt to resolve this ambiguity has involved the introduction of additional degrees of freedom whose purpose is to essentially track the entangling surface's location in spacetime~\cite{Donnelly:2016auv,Donnelly:2016rvo,Speranza:2017gxd}.

It will be shown in the course of this paper that once one has actually calculated the reduced density matrix, this ambiguity goes away. The reduction procedure imposes certain smoothness conditions on the metric at the entangling surface, and those conditions break diffeomorphism invariance. One ends up with a density matrix whose elements are weighted by a factor which depends on the location of the entangling surface, and this factor becomes sharply peaked in the semiclassical limit. In a sense, the reduction procedure has implicitly introduced degrees of freedom of the kind espoused in~\cite{Donnelly:2016auv,Donnelly:2016rvo,Speranza:2017gxd}.

The rest of the paper proceeds as follows. In Section \ref{Section: Density matrices}, we describe the class of density matrices that we are interested in, and explain what happens when we calculate the reduced density matrix. Section \ref{Section: Hamiltonian} contains a description of the Hamiltonian dynamics in the region near the entangling surface, making use of the covariant phase space method~\citecovariant. It includes a resolution of the ambiguities inherent in that approach~\citeambiguities, to the extent that they will affect our results. Of primary interest is the Hamiltonian charge which generates boosts about the entangling surface, and in Section \ref{Section: Entangling surface charge} we compute this charge, showing that it is proportional to the area in general relativity, and a dynamical generalisation of the Wald entropy in higher derivative theories of gravity. We take the semiclassical limit in Section \ref{Section: Semiclassical limit}, and explain how this naturally leads to domination by states for which the entangling surface boost charge is minimised. In Section \ref{Section: entropy}, we calculate the contribution to the entropy of the reduced density matrix originating at the entangling surface in the semiclassical limit. In particular we show agreement with the Ryu-Takayanagi conjecture. Finally, we conclude the paper in Section \ref{Section: Discussion}, summarising our results, and suggesting future directions.

\section{Density matrices and effective spacetimes}
\label{Section: Density matrices}

We work in $D$ spacetime dimensions. Let $\mathcal{H}$ be a Hilbert space for the canonical quantization of pure GR\footnote{More generally, we can replace $\mathcal{H}$ and $I$ in \eqref{Equation: Einstein-Hilbert} by the Hilbert space and action of whichever theory of gravity we wish to analyse.} on a spatial $(D-1)$-surface $\Sigma$, and let $\{\ket{\gamma}\}$ be an eigenbasis in $\mathcal{H}$ of the $(D-1)$-metric $\gamma_{ij}$.

We will consider in this paper a class of density matrices $\rho$ whose elements in the basis $\{\ket{\gamma}\}$ can be approximately written as Euclidean path integrals
\begin{equation}
	\mel*{\gamma^1}{\rho}{\gamma^0} = \int^{\gamma^1}_{\gamma^0}\Dd{g} \exp(-\frac1\hbar I),
	\label{Equation: density matrix elements}
\end{equation}
where
\begin{equation}
	I[g] = \frac1{16\pi G}\int_{\mathcal{M}}\dd[D]{x}\sqrt{\det g}R + \frac1{8\pi G} \int_{\partial\mathcal{M}} \dd[D-1]{x} \sqrt{\det\gamma}K.
	\label{Equation: Einstein-Hilbert}
\end{equation}
is the Einstein-Hilbert action for a $D$-metric $g$ on a $D$-dimensional manifold $\mathcal{M}$, which we will call the `effective spacetime'.\footnote{To avoid confusion, we will use the term `spacetime' instead of just `space', even though this is a Euclidean path integral.} In this expression, $R$ is the scalar curvature of $g$, $\gamma$ is the induced metric on the boundary of $\mathcal{M}$, and $K$ is the extrinsic curvature of that boundary, which is taken to consist of three parts: a `past' boundary $\Sigma^0$, a `future' boundary $\Sigma^1$, and a `spatial' boundary  $\mathcal{B}$. As is well-known, the boundary term is required for the variational principle to be well-defined~\cite{GibbonsHawkingPhysRevD.15.2752}. Topologically, we assume that $\Sigma^1,\Sigma^2$ are homeomorphic to $\Sigma$, which we notate $\Sigma^0\sim \Sigma^1\sim\Sigma$. The range of the path integral includes all smooth Euclidean metrics for which $\gamma|_{\Sigma^0}=\gamma^0$, $\gamma|_{\Sigma^1}=\gamma^1$, and for which $\gamma$ obeys certain boundary conditions at $\mathcal{B}$. The topology of $\mathcal{M}$ and boundary conditions at $\mathcal{B}$ are dependent on the density matrix under consideration. 

We can represent the elements of such density matrices diagrammatically in the following way.
\begin{equation}
	\mel*{\gamma^1}{\rho}{\gamma^0} :\qquad
	\begin{tikzpicture}[thick,baseline={([yshift=-.5ex]current bounding box.center)},scale=0.8]
		\fill[blue, opacity=0.15] (0,0) -- (8,0) -- (8,-4) -- (0,-4) -- cycle;
		\draw[red] (0,0) -- (0,-4);
		\node[right] at (0,-2) {$\mathcal{B}$};
		\draw[red] (8,0) -- (8,-4);
		\node[left] at (8,-2) {$\mathcal{B}$};
		\draw[red,dashed] (0,0) -- (8,0) node[above, midway] {$\gamma^1$};
		\node[below] at (4,0) {$\Sigma^1$};
		\draw[red,dashed] (0,-4) -- (8,-4) node[below, midway] {$\gamma^0$};
		\node[above] at (4,-4) {$\Sigma^0$};
		\node[scale=1.5] at (4,-2) {$\mathcal{M}$};
	\end{tikzpicture}
	\label{Diagram: generic}
\end{equation}
This figure represents the $D$-manifold $\mathcal{M}$ over which the action is to be evaluated. Each point in the figure corresponds to a $(D-2)$-dimensional surface in $\mathcal{M}$. The dashed lines are the past and future boundaries on which the induced metric is determined by which matrix element interests us. The solid lines are the spatial boundary where the induced metric is determined by the boundary conditions.

Included in this class of density matrices are thermal density matrices $\rho=e^{-\frac1\hbar\beta H}$, where $H$ is the Hamiltonian, and $\beta$ is the inverse temperature (we are explicitly including the factor of $\hbar$ so that we can take the semiclassical limit $\hbar\to 0$ at a later stage). In this case, $\mathcal{M}$ can be smoothly foliated by a family of surfaces $\Sigma_\tau\sim\Sigma$, $\tau\in [0,\beta]$, and the imaginary Hamiltonian $iH$ can be viewed as the infinitesimal generator which moves between these surfaces at a speed of one, as measured in units of imaginary time $\tau$. In what is perhaps an abuse of language, we will refer to the $\Sigma_t$ as Cauchy surfaces. We identify $\Sigma_0 = \Sigma^0$, $\Sigma_\beta=\Sigma^1$, and $\mathcal{B} = \bigcup_\tau \partial\Sigma_\tau$.
\begin{equation}
	\mel*{\gamma^1}{e^{-\frac1\hbar\beta H}}{\gamma^0} : \qquad
	\begin{tikzpicture}[thick,baseline={([yshift=-.5ex]current bounding box.center)},scale=0.8]
		\fill[blue, opacity=0.15] (0,0) -- (8,0) -- (8,-4) -- (0,-4) -- cycle;
		\draw[red] (0,0) -- (0,-4);
		\draw[red] (8,0) -- (8,-4);
		\draw[red,dashed] (0,0) -- (8,0) node[above, midway] {$\gamma^1$};
		\draw[red,dashed] (0,-4) -- (8,-4) node[below, midway] {$\gamma^0$};
		
		\foreach \y in {-0.25,-0.5,...,-1.25} {
			\draw[blue,opacity=0.5] (0,\y) -- (8,\y);
		}
		\foreach \y in {-1.5,-1.75,...,-2.5} {
			\draw[blue,opacity=0.5] (0,\y) -- (3.5,\y);
			\draw[blue,opacity=0.5] (4.7,\y) -- (8,\y);
		}
		\foreach \y in {-2.75,-3,...,-3.75} {
			\draw[blue,opacity=0.5] (0,\y) -- (8,\y);
		}
	
		\draw[decorate,decoration={brace,amplitude=5pt}] (3.6,-1.4) -- (3.6,-2.6);
		\node[right] at (3.75,-2) {$\Sigma_\tau$};
		
		\draw[->] (8.15,-2) -- (8.15,-1.75) node[right,midway] {$iH$};
		\node[right] at (8,0) {$\tau=\beta$};
		\node[right] at (8,-4) {$\tau=0$};
	\end{tikzpicture}
	\label{Diagram: thermal}
\end{equation}
Conversely, if for a given density matrix $\rho$ we can find a foliation with these properties, and a Hamiltonian $H$ for which $iH$ infinitesimally generates evolution from one surface to the next, we can interpret $\rho = e^{-\frac1\hbar\beta H}$ as being thermal with respect to $H$. The inverse temperature $\beta$ is determined by the boundary conditions implicit in $\rho$.

Thinking about density matrices in this way can be useful for conceptualising various calculations. For example, the partition function $\mathcal{Z}$ is defined as the trace of the density matrix. One can calculate $\mathcal{Z}$ by setting $\gamma^1=\gamma^0$ in the matrix element, and integrating over all $\gamma^0$. The result can then be interpreted as a path integral over the $D$-manifold obtained from $\mathcal{M}$ by identifying $\Sigma^1$ with $\Sigma^0$. In the case of a thermal density matrix, this corresponds to making the imaginary time direction periodic. Diagrammatically, one can imagine deforming $\mathcal{M}$ so that the past and future boundaries meet. Because the path integral is diffeomorphism invariant, such deformations are allowed, so long as they do not change the induced boundary metric.
\begin{equation}
	\mathcal{Z} = \int\Dd{\gamma}\mel*{\gamma}{\rho}{\gamma} :\qquad
	\begin{tikzpicture}[thick,baseline={([yshift=-.5ex]current bounding box.center)}]
		\fill[blue, opacity=0.15] (0,0) -- (4,0) arc (90:20:0.3 and 1) coordinate (a1) -- +(-4.15,0) coordinate (a2) arc (20:90:0.15 and 1);
		\draw[red,dashed] (a1) -- (a2);
		\fill[blue, opacity=0.15] (0,-2) -- (4,-2) arc (-90:-20:0.3 and 1) coordinate (b1) -- +(-4.15,0) coordinate (b2) arc (-20:-90:0.15 and 1);
		\draw[red,dashed] (b1) -- (b2);
		\fill[blue, opacity=0.1] (0,0) -- (4,0) arc (90:270:0.3 and 1) -- (0,-2) arc (270:90:0.15 and 1);
		\draw[blue,opacity=0.3] (0,0) -- (4,0);
		\draw[blue,opacity=0.3] (0,-2) -- (4,-2);
		\draw[red] (a2) arc (20:340:0.15 and 1);
		\draw[red] (a1) arc (20:90:0.3 and 1);
		\draw[red] (b1) arc (-20:-90:0.3 and 1);
		\draw[red,dotted,opacity=0.5] (4,0) arc (90:270:0.3 and 1);
		
		\node[red] at (2.3,-1) {$\gamma^1=\gamma^0$};
		\node[above] at (2.3,-0.7) {$\Sigma^1$};
		\node[below] at (2.3,-1.3) {$\Sigma^0$};
		
		\node at (5,-1) {$\longrightarrow$};
		
		\begin{scope}[shift={(6,0)}]
			\fill[blue, opacity=0.15] (0,0) -- (4,0) arc (90:-90:0.3 and 1) -- (0,-2) arc (-90:90:0.15 and 1);
			\fill[blue, opacity=0.1] (0,0) -- (4,0) arc (90:270:0.3 and 1) -- (0,-2) arc (270:90:0.15 and 1);
			\draw[blue,opacity=0.3] (0,0) -- (4,0);
			\draw[blue,opacity=0.3] (0,-2) -- (4,-2);
			\draw[blue,opacity=0.3,dashed] (0.15,-1) -- (4.3,-1);
			\draw[red] (0,0) arc (90:450:0.15 and 1);
			\draw[red] (4,0) arc (90:-90:0.3 and 1);
			\draw[red,dotted,opacity=0.5] (4,0) arc (90:270:0.3 and 1);
		\end{scope}
	\end{tikzpicture}
\end{equation}

\subsection{Reduced density matrix}

Now suppose $A$ is a subregion of $\Sigma$, and let $\bar{A}$ be the complement of $A$ in $\Sigma$. The reduced density matrix $\rho_A$ is obtained by taking $\rho$ and tracing over all degrees of freedom present in $\bar{A}$. One can calculate reduced matrix elements $\mel*{\gamma^1_A}{\rho_A}{\gamma^0_A}$ by setting $\gamma^1|_{\bar{A}}=\gamma^0|_{\bar{A}}$ in the matrix element of $\rho$, and integrating over all $\gamma^0|_{\bar{A}}$. The result can be interpreted as a path integral over the $D$-manifold obtained from $\mathcal{M}$ by identifying $\bar{A}^1$ with $\bar{A}^0$, where $\bar{A}^1,\bar{A}^0$ are the subregions of $\Sigma^1,\Sigma^0$ respectively which correspond to $\bar{A}$ in the topological identifications $\Sigma^1,\Sigma^0\sim\Sigma$. This new effective spacetime appears in the following diagram.
\begin{equation}
	\mel*{\gamma_A^1}{\rho_A}{\gamma_A^0} = \int\Dd{\gamma_{\bar{A}}}\mel*{\gamma_A^1,\gamma_{\bar{A}}}{\rho}{\gamma_A^0,\gamma_{\bar{A}}} : \qquad
	\begin{tikzpicture}[thick,baseline={([yshift=-.5ex]current bounding box.center)}]
		\fill[blue, opacity=0.1] (2,0) -- (4,0) .. controls (5,0) and (6,0) .. (8,1) arc (90:270:0.3 and 2) .. controls (6,-2) and (5,-2) .. (4,-2) -- (2,-2) arc (270:90:0.15 and 1);
		\draw[blue,opacity=0.3] (2,0) -- (4,0) .. controls (5,0) and (6,0) .. (8,1);
		\draw[blue,opacity=0.3] (8,-3) .. controls (6,-2) and (5,-2) .. (4,-2) -- (2,-2);
		\draw[blue,opacity=0.3,dashed] (2.15,-1) -- (4.3,-1);
		\draw[red] (2,0) arc (90:450:0.15 and 1);
		\draw[red] (8,1) arc (90:50:0.3 and 2) coordinate (a);
		\draw[red] (8,-3) arc (-90:-50:0.3 and 2) coordinate (b);
		\draw[red,dotted,opacity=0.5] (8,1) arc (90:270:0.3 and 2);
		
		\path (a) -- +(-0.2,-0.2) coordinate (ac);
		\path (b) -- +(-0.2,0.2) coordinate (bc);
		
		\fill[blue, opacity=0.15] (2,0) -- (4,0) .. controls (5,0) and (6,0) .. (8,1) arc (90:50:0.3 and 2) .. controls (ac) and (6,-1) .. (4.3,-1) .. controls (6,-1) and (bc) .. (b) arc (-50:-90:0.3 and 2) .. controls (6,-2) and (5,-2) .. (4,-2) -- (2,-2) arc (-90:90:0.15 and 1);
		
		\draw[red, dashed] (a) .. controls (ac) and (6,-1) .. (4.3,-1);
		\draw[red, dashed] (b) .. controls (bc) and (6,-1) .. (4.3,-1);
		
		\node[right,red] at (6.8,-0.6) {$\gamma^1_A$};
		\node[right,red] at (6.8,-1.4) {$\gamma^0_A$};
		
		\node[left] at (6.5,-0.3) {$A^1$};
		\node[left] at (6.5,-1.7) {$A^0$};
		
		\fill[red] (4.3,-1) circle (0.07);
		\node[above] at (4.3,-1) {$\Upsilon$};
		
		\draw[decorate,decoration={brace,amplitude=5pt}] (4,-2.1) -- (2,-2.1);
		\node[below] at (3,-2.3) {$\bar{A}$};
	\end{tikzpicture}
\end{equation}
The dot labelled $\Upsilon$ represents the $(D-2)$-surface in $\Sigma$ dividing the two regions $A$ and $\bar{A}$. This surface is commonly called the `entangling' surface.

It is useful to deform this diagram so that it becomes planar. To do so, imagine taking the component of $\mathcal{B}$ on the left of the diagram, and pushing it into the tube so that it emerges on the right side of $\Upsilon$. Unrolling $A^1$ upwards and $A^0$ downwards, the result is:
\begin{equation}
	\mel*{\gamma_A^1}{\rho_A}{\gamma_A^0} :\qquad
	\begin{tikzpicture}[scale=1.1,thick,baseline={([yshift=-.5ex]current bounding box.center)}]
		\fill[blue, opacity=0.15] (7,0) 
		.. controls (1,0) and (0,0) .. (0,0.75)
		.. controls (0,1.5) and (1,1.5) .. (1.5,1.5)
		.. controls (1,1.5) and (0,1.5) .. (0,2.25)
		.. controls (0,3) and (1,3) .. (7,3);
	
		\draw[red, dashed] (7,0) 
			.. controls (1,0) and (0,0) .. (0,0.75)
			.. controls (0,1.5) and (1,1.5) .. (1.5,1.5)
			.. controls (1,1.5) and (0,1.5) .. (0,2.25)
			.. controls (0,3) and (1,3) .. (7,3);
		\draw[red] (7,0) -- (7,3);
		\draw[red,fill=white] (4.5,1.5) circle (0.7);
		\node[right] at (5.2,1.5) {$\mathcal{B}_{\bar{A}}$};
		\node[left] at (7,1.5) {$\mathcal{B}_A$};
		\fill[red] (1.5,1.5) circle (0.07);
		\node[right] at (1.5,1.5) {$\Upsilon$};
		\node[below] at (3.5,3) {$A^1$};
		\node[above] at (3.5,0) {$A^0$};
		\node[above,red] at (3.5,3) {$\gamma^1_A$};
		\node[below,red] at (3.5,0) {$\gamma^0_A$};
	\end{tikzpicture}
	\label{Diagram: reduced}
\end{equation}
Thus the effective spacetime for a reduced density matrix looks much like the generic one in \eqref{Diagram: generic}, but with some additional distinguishing features, namely the replacement of part of the spatial boundary by the entangling surface $\Upsilon$, as well as the addition of a new interior component for the spatial boundary, which we will refer to as $\mathcal{B}_{\bar{A}}$, since it is the part of $\mathcal{B}$ adjoined to $\bar{A}$. The rest of the spatial boundary will be referred to as $\mathcal{B}_A$.

We have drawn the diagram near $\Upsilon$ in the above way\footnote{The reader might wonder why we have chosen capital upsilon $\Upsilon$ to represent the entangling surface. It is because the shape of this letter closely resembles the spacetime near $\Upsilon$.} in order to emphasise the following point. If we assume that the $D$-metric $g$ is smooth, and in particular free of conical defects near $\Upsilon$, and that the surfaces $\Sigma^1$ and $\Sigma^0$ are smoothly embedded in the effective spacetime, then the angle subtended by a path from a point on $A^0$ near $\Upsilon$ to a point on $A^1$ near $\Upsilon$ must be close to $2\pi$. Thus in order for this flattened diagram to be faithful to these smoothness properties, the angle between $A^1$ and $A^0$ at $\Upsilon$ must be $2\pi$.

\subsection{Foliation of reduced effective spacetime}

The existence of the interior component of the spatial boundary generically forbids a foliation of this effective spacetime of the type necessary for $\rho_A$ to be interpreted as a thermal density matrix as in \eqref{Diagram: thermal}. However, we can get something close. Consider first just the region near $\Upsilon$. We can foliate this region by a set of surfaces $A_\tau$, $\tau\in[0,2\pi]$, as in the following diagram.
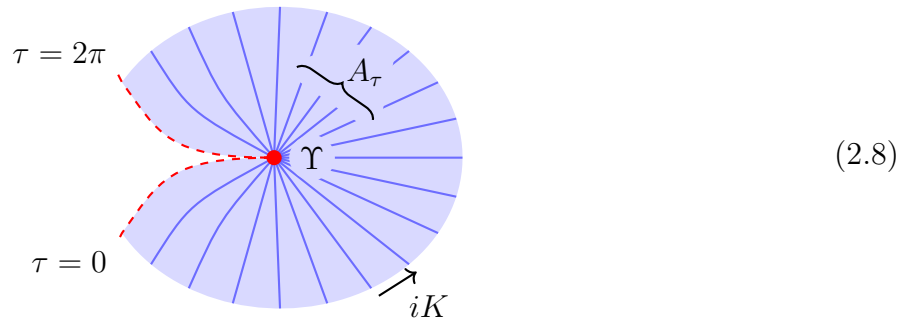
\begin{equation}
	\begin{tikzpicture}[thick,baseline={([yshift=-.5ex]current bounding box.center)}]
		\fill[blue!15] (0,1) 
			.. controls (0.5,0.3) and (0.6,0) .. (2,0)
			.. controls (0.6,0) and (0.5,-0.3) .. (0,-1)
			arc (-150:150:2.4 and 2) 
				coordinate [pos=0.05] (c1)
				coordinate [pos=0.1] (c2)
				coordinate [pos=0.15] (c3)
				coordinate [pos=0.2] (c4)
				coordinate [pos=0.25] (c5)
				coordinate [pos=0.3] (c6)
				coordinate [pos=0.35] (C6)
				coordinate [pos=0.4] (c7)
				coordinate [pos=0.45] (c8)
				coordinate [pos=0.5] (c9)
				coordinate [pos=0.55] (c10)
				coordinate [pos=0.6] (c11)
				coordinate [pos=0.65] (c12)
				coordinate [pos=0.7] (c13)
				coordinate [pos=0.75] (c14)
				coordinate [pos=0.8] (c15)
				coordinate [pos=0.85] (c16)
				coordinate [pos=0.9] (c17)
				coordinate [pos=0.95] (c18);
		\draw[blue,opacity=0.5] (c1) .. controls +(0.5,0.8) .. (2,0);
		\draw[blue,opacity=0.5] (c2) .. controls +(0.4,0.9) .. (2,0);
		\draw[blue,opacity=0.5] (c3) -- (2,0);
		\draw[blue,opacity=0.5] (c4) -- (2,0);
		\draw[blue,opacity=0.5] (c5) -- (2,0);
		\draw[blue,opacity=0.5] (c6) -- (2,0);
		\draw[blue,opacity=0.5] (C6) -- (2,0);
		\draw[blue,opacity=0.5] (c7) -- (2,0);
		\draw[blue,opacity=0.5] (c8) -- (2,0);
		\draw[blue,opacity=0.5] (c9) -- (2,0);
		\draw[blue,opacity=0.5] (c10) -- (2,0);
		\draw[blue,opacity=0.5] (c11) -- (2,0);
		\draw[blue,opacity=0.5] (c12) -- (2,0);
		\draw[blue,opacity=0.5] (c13) -- (2,0);
		\draw[blue,opacity=0.5] (c14) -- (2,0);
		\draw[blue,opacity=0.5] (c15) -- (2,0);
		\draw[blue,opacity=0.5] (c16) -- (2,0);
		\draw[blue,opacity=0.5] (c17) .. controls +(0.4,-0.9) .. (2,0);
		\draw[blue,opacity=0.5] (c18) .. controls +(0.5,-0.8) .. (2,0);
		
		\draw[red,dashed] (-0.05,1.1) -- (0,1) 
			.. controls (0.5,0.3) and (0.6,0) .. (2,0)
			.. controls (0.6,0) and (0.5,-0.3) .. (0,-1)
			-- (-0.05,-1.1);
		\fill[red] (2,0) circle (0.1);
		\fill[blue!15] (2.5,0) circle (0.3);
		\node at (2.5,0) {$\Upsilon$};
		
		\node[above left] at (-0.05,1.1) {$\tau=2\pi$};
		\node[below left] at (-0.05,-1.1) {$\tau=0$};
		
		\path (c6) -- +(0.1,-0.1) coordinate (A);
		\path (C6) -- +(0.1,-0.1) coordinate (B);
		
		\draw[->] (A) -- (B) node[midway,below right] {$iK$};
		
		\fill[blue!15] (2.3,1.3) -- (3.1,1.5) -- (3.6,1.4) -- (3.4,0.5) -- (3.2,0.4) -- (2.2,1.1) -- cycle;
		
		\draw[decorate,decoration={brace,amplitude=5pt}] (2.3,1.2) -- (3.3,0.5);
		\node[above right] at (2.8,0.8) {$A_\tau$};
	\end{tikzpicture}
\end{equation}
$A_0$ is identified with the part of $A^0$ near $\Upsilon$, and $A_{2\pi}$ is identified with the part of $A^1$ near $\Upsilon$. The parameter $\tau$ measures the angle at which the surface $A_\tau$ meets $\Upsilon$. The imaginary Hamiltonian $iK$ in this instance generates rotations about $\Upsilon$, and the Hamiltonian $K$ generates an $\Upsilon$-preserving boost in Lorentzian spacetime. The inverse temperature is $\beta=2\pi$. Therefore, at least in a sense local to $\Upsilon$, the reduced density matrix is thermal with respect to a boost charge evaluated near $\Upsilon$, at temperature $1/2\pi$. Imaginary angle $\eta = i\tau$ is rapidity. This is a gravitational generalisation of a well-known result for the density matrix of a half space~\cite{Bisognano-Wichmann}.

There are many equivalent ways in which one could now extend this foliation to get an idea of the form of the full reduced density matrix. One such possibility is described next, but keep in mind that the result of this paper should be independent of this procedure. Because of the diffeomorphism invariance properties of the path integral, all that is really required for the analysis to follow through is that the evolution is thermal near $\Upsilon$ in the way just described.

We assume that $\mathcal{B}_A$ has the topology of $\Upsilon\times[0,2\pi]$. Then we can extend the definition of each $A_\tau$ so that it meets $\mathcal{B}_A$ at a unique cross section. Together with the condition that $A_\tau$ lies above $\mathcal{B}_{\bar{A}}$ for $\tau\ge\pi$, and below $\mathcal{B}_{\bar{A}}$ for $\tau<\pi$ (where `above' and `below' are with respect to \eqref{Diagram: reduced}), this enables us to completely foliate the effective spacetime by the surfaces $A_\tau$. This is illustrated below.
\begin{equation}
	\mel*{\gamma_A^1}{\rho_A}{\gamma_A^0} :\qquad
	\begin{tikzpicture}[scale=1.1,thick,baseline={([yshift=-.5ex]current bounding box.center)}]
		\fill[blue, opacity=0.15] (7,0) 
		.. controls (1,0) and (0,0) .. (0,0.75)
		.. controls (0,1.5) and (1,1.5) .. (1.5,1.5)
		.. controls (1,1.5) and (0,1.5) .. (0,2.25)
		.. controls (0,3) and (1,3) .. (7,3);
		
		\begin{scope}[blue,opacity=0.5]
			\path (4.5,2.25) arc (90:170:0.75) coordinate (a1);
			\path (4.5,2.4) arc (90:140:0.9) coordinate (a2);
			\path (4.5,2.55) arc (90:115:1.05) coordinate (a3);
			\draw (a1) arc (170:10:0.75) coordinate (b1);
			\draw (a2) arc (140:40:0.9) coordinate (b2);
			\draw (a3) arc (115:65:1.05) coordinate (b3);
			\draw (b1) .. controls (5.35,1.55) .. (7,1.55);
			\draw (b2) .. controls (5.45,1.9) .. (7,1.9);
			\draw (b3) .. controls (5.3,2.3) .. (7,2.3);
			\draw (7,2.7) -- (1.75,2.7)
				.. controls (0,2.7) and (0,1.7) .. (1.5,1.5);
			\draw (1.5,1.5) .. controls (3.65,1.5) .. (a1);
			\draw (1.5,1.5) .. controls (1.65,1.8) and (3.6,1.7) .. (a2);
			\draw (1.5,1.5) .. controls (0.7,2.5) and (3.4,2.1) .. (a3);
		\end{scope}

		\begin{scope}[blue,opacity=0.5,yscale=-1,shift={(0,-3)}]
			\path (4.5,2.25) arc (90:170:0.75) coordinate (a1);
			\path (4.5,2.4) arc (90:140:0.9) coordinate (a2);
			\path (4.5,2.55) arc (90:115:1.05) coordinate (a3);
			\draw (a1) arc (170:10:0.75) coordinate (b1);
			\draw (a2) arc (140:40:0.9) coordinate (b2);
			\draw (a3) arc (115:65:1.05) coordinate (b3);
			\draw (b1) .. controls (5.35,1.55) .. (7,1.55);
			\draw (b2) .. controls (5.45,1.9) .. (7,1.9);
			\draw (b3) .. controls (5.3,2.3) .. (7,2.3);
			\draw (7,2.7) -- (1.75,2.7)
			.. controls (0,2.7) and (0,1.7) .. (1.5,1.5);
			\draw (1.5,1.5) .. controls (3.65,1.5) .. (a1);
			\draw (1.5,1.5) .. controls (1.65,1.8) and (3.6,1.7) .. (a2);
			\draw (1.5,1.5) .. controls (0.7,2.5) and (3.4,2.1) .. (a3);
		\end{scope}	
			
		\draw[red, dashed] (7,0) 
		.. controls (1,0) and (0,0) .. (0,0.75)
		.. controls (0,1.5) and (1,1.5) .. (1.5,1.5)
		.. controls (1,1.5) and (0,1.5) .. (0,2.25)
		.. controls (0,3) and (1,3) .. (7,3);
		\draw[red] (7,0) -- (7,3);
		\draw[red,fill=white] (4.5,1.5) circle (0.7);
		\fill[red] (1.5,1.5) circle (0.07);
		\node[above,red] at (3.5,3) {$\gamma^1_A$};
		\node[below,red] at (3.5,0) {$\gamma^0_A$};
		
		\node[right] at (7,0) {$\tau=0$};
		\node[right] at (7,1.5) {$\tau=\pi$};
		\node[right] at (7,3) {$\tau=2\pi$};
	\end{tikzpicture}
	\label{Diagram: reduced foliation}
\end{equation}
So this is almost thermal, but the obvious caveat is that the transition from $\tau<\pi$ to $\tau\ge \pi$ is not a smooth one. Let $\mathscr{U}$ be an operator which goes from the surface $A_\tau$ for $\tau$ just below $\pi$, to the one for $\theta$ just above $\pi$. The exact form of $\mathscr{U}$ is determined by the boundary conditions implicit in the original density matrix $\rho$. We might interpret $\mathscr{U}$ as an operator which inserts the interior boundary into the state at $\tau=\pi$. More generally, $\mathscr{U}$ accounts for any topologically non-trivial evolution that happens away from the entangling surface.

In summary, the evolutions in the ranges $\tau<\pi$ and $\tau>\pi$ are thermal, and the full evolution consists of evolution through $0\le\tau<\pi$, then an application of $\mathscr{U}$, then evolution through $\pi\le\tau\le2\pi$. We can thus write the elements of the reduced density matrix as
\begin{equation}
	\mel*{\gamma^1_A}{\rho_A}{\gamma^0_A} = \mel*{\gamma^1_A}{e^{-\frac1\hbar\pi K} \mathscr{U} e^{-\frac1\hbar\pi K}}{\gamma^0_A},
	\label{Equation: reduced density decomposition}
\end{equation}
where we have extended the definition of $iK$ so that it is the infinitesimal generator of evolution along the leaves of the foliation in \eqref{Diagram: reduced foliation}. Examining \eqref{Equation: reduced density decomposition}, one finds that the matrix elements of $\rho_A$ in the basis $\{\ket{\gamma_A}\}$ are equivalent to the matrix elements of $\mathscr{U}$ in the non-unitarily transformed basis $\{e^{-\frac1\hbar\pi K}\ket{\gamma_A}\}$. 

\section{Hamiltonian dynamics near the entangling surface}
\label{Section: Hamiltonian}

In this section we will develop the formalism necessary to properly analyse the operator $K$. We will need an appropriate Hamiltonian description of the dynamics near $\Upsilon$, and we find it most convenient to use the covariant phase space method~\citecovariant. Normally there are boundary ambiguities in such a formulation~\citeambiguities. We will show how to resolve these ambiguities at the entangling surface, where they are relevant to our purposes.

\subsection{Covariant phase space method}

Let $*$ be the spacetime Hodge dual operator. The local dynamics of a covariant field theory in $D$ dimensions is described by a $D$-form $*L$ known as the Lagrangian density. We will use $\phi$ to denote all the fields in the theory, including the metric. $L$ is a local function of $\phi$.\footnote{$L$ can of course depend on derivatives of $\phi$. Whenever we speak of a `local function of $X$' in this paper, we mean a function that depends locally on $X$ and its spacetime derivatives $\nabla X,\nabla\nabla X,\dots$.} Under an arbitrary infinitesimal variation of the fields $\phi\to\phi+\delta \phi$, the change in the Lagrangian density can be written
\begin{equation}
	\delta(*L) = E \cdot \delta\phi + \dd{\theta}.
	\label{Equation: Lagrangian density variation}
\end{equation}
In this expression $E=\fdv{(*L)}{\phi}$ is the Euler-Lagrange derivative of $*L$ with respect to $\phi$, and the dot denotes a sum over all of the components of the fields. If for a particular field configuration the equations of motion $E=0$ hold, that field configuration is said to be on-shell. The covariant phase space is the space of all on-shell field configurations.

The $(D-1)$-form $\theta$ is a local function of $\phi$ and a linear local function of $\delta\phi$, and is called the symplectic potential density.\footnote{Strictly speaking it is the \emph{pre}symplectic potential density, because we have not yet carried out gauge reduction. This will not be important in this paper, so to avoid confusion we will just use the term `symplectic' instead of `presymplectic' wherever applicable.} We obtain the symplectic potential $\Theta$ by integrating $\theta$ over a Cauchy surface $\Sigma$.
\begin{equation}
	\Theta[\phi,\delta\phi] = \int_\Sigma \theta(\phi,\delta\phi).
\end{equation}
The field variation $\phi\to\phi+\delta\phi$, can be viewed as a vector field in field space. Hence $\Theta$, being a linear functional of $\delta\phi$, is a 1-form in field space. The symplectic form $\Omega$ is a 2-form on field space which one obtains by taking the field space exterior derivative of $\Theta$. It can be written
\begin{equation}
	\Omega[\phi,\delta_1\phi,\delta_2\phi] = \int_\Sigma\omega(\phi,\delta_1\phi,\delta_2\phi), 
\end{equation}
where
\begin{equation}
	\omega(\phi,\delta_1\phi,\delta_2\phi) = \delta_1\theta(\phi,\delta_2\phi) - \delta_2\theta(\phi,\delta_1\phi) - \theta(\phi,\delta_{12}\phi).
	\label{Equation: symplectic form density}
\end{equation}
In this equation $\delta X$ means the change in $X$ resulting from the variation $\phi\to\phi+\delta\phi$, and $\delta_{12} = [\delta_1,\delta_2]$ is the commutator of two variations $\delta_1$ and $\delta_2$. (i.e.\ their Lie bracket when viewed as vector fields on field space).

\subsection{Fixing ambiguities}

The formalism described in the previous subsection suffers from two ambiguities. First, the local dynamics are contained within the equations of motion, and these do not change if we modify the Lagrangian density by the addition of an exact $D$-form, $L\to L+\dd{\mu}$. The symplectic potential correspondingly changes by $\Theta\to\Theta + \delta \qty(\int_\Sigma\mu)$, but the symplectic form $\Omega$ is clearly invariant. In other words this change corresponds to a canonical transformation. Since the physically relevant information is contained in $\Omega$, we do not need to be concerned with this ambiguity.

The second ambiguity is more serious. The equation \eqref{Equation: Lagrangian density variation} only specifies the symplectic potential up to the addition of a closed $(D-1)$-form that is linearly locally dependent on $\delta\phi$. Any such closed $(D-1)$-form is exact~\cite{Wald:2010:doi:10.1063/1.528839}, so this ambiguity is of the form $\theta \to \theta + \dd(Y(\phi,\delta\phi))$. Under such an addition the symplectic form genuinely does change, and so there are physical consequences. To be precise it changes by the addition of a boundary term:
\begin{equation}
	\Omega[\phi,\delta_1\phi,\delta_2\phi] \to \Omega[\phi,\delta_1\phi,\delta_2\phi] + \int_{\partial\Sigma} \delta_1Y(\phi,\delta_2\phi) - \delta_2Y(\phi,\delta_1\phi) - Y(\phi,\delta_{12}\phi).
\end{equation}
Thus such a modification affects the boundary degrees of freedom. 

The reason for this ambiguity is that one has failed to specify what exactly goes on at the boundary $\partial \Sigma$. Without such a specification, the theory we are concerned with is ill-defined. In our case, we know exactly what goes on at the entangling surface, where time evolution just consists of a rotation around $\Upsilon$. We should therefore be able to fix this ambiguity at $\Upsilon$.

To understand how this will work, it is instructive to recall how one defines the symplectic potential in classical mechanics, where boundary ambiguities manifestly do not exist. Consider a theory of an evolving degree of freedom $q$. The action for evolution between the times $t=t_0$ and $t=t_1$ is given by
\begin{equation}
	S = \int^{t_1}_{t_0} \dd{t} L(q,\dot{q}).
\end{equation}
The variation of the action is 
\begin{equation}
	\delta S = \int^{t_1}_{t_0} \dd{t} \fdv{L}{q}\delta q + \qty[\pdv{L}{\dot q}\delta q]_{t_0}^{t_1},
\end{equation}
where $\fdv{L}{q}=\pdv{L}{q}-\dv{t}\pdv{L}{\dot{q}}$ is the Euler-Lagrange derivative of $L$ with respect to $q$. On-shell $\fdv{L}{q}$ vanishes, and we can write $\delta S = \Theta(t_1)-\Theta(t_0)$, where 
\begin{equation}
	\Theta(t) = \qty[\pdv{L}{\dot q}\delta q]_t + C[q,\delta q_0].
	\label{Equation: mechanics symplectic potential}
\end{equation}
In this expression, $C[q,\delta q_0]$ is an unspecified field space function which is independent of $t$ and linear in $\delta q$, and we have written $\delta q_0$ to indicate that $C$ depends on the zero mode of $\delta q$, i.e.\ its time-independent part.\footnote{$C$ can \emph{only} depend on $\delta q$ through its zero mode. This follows from the fact that $\Theta(t)$ must be linear in $\delta q$, and that it must be of the form \eqref{Equation: mechanics symplectic potential} for any arbitrary choice of $\delta q$.} This almost completely defines the symplectic potential $\Theta$, with the only remaining ambiguity in the choice of $C$. Note that our goal will be to calculate the generator of time evolution, and the choice of $C$ will not affect that calculation.

Now we try this argument again, but from the point of view of field theory near $\Upsilon$. We will initially regulate the region near $\Upsilon$ by removing a disk $\mathcal{D}_\epsilon$ in the normal plane to the entangling surface centered at $\Upsilon$ and of radius $\epsilon$. At the end we will take the limit $\epsilon\to 0$. We consider the action for evolution between the angles $\tau=\tau_0$ and $\tau=\tau_1$; we denote the relevant region in spacetime by $\mathcal{M}(\tau_0,\tau_1)$. The surface at angle $\tau$ we label $A_\tau$. The initial and final surfaces are therefore $A_{\tau_0},A_{\tau_1}$ respectively. The part of $\partial\mathcal{D}_\epsilon$ between $\tau=\tau_0$ and $\tau=\tau_1$ we label $B_\epsilon$. This region of spacetime is depicted below.
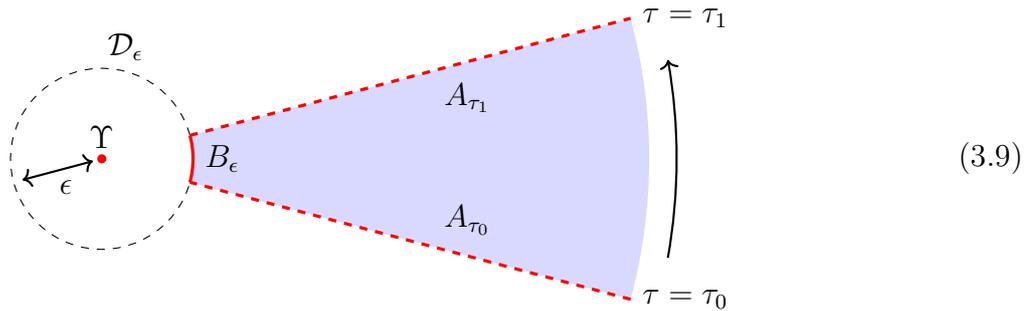
\begin{equation}
	\begin{tikzpicture}[baseline={([yshift=-.5ex]current bounding box.center)},rotate=-15,scale=1.2]
		\fill[blue,opacity=0.15] (6,0) -- (1,0) arc (0:30:1) -- ++(30:5) arc (30:0:6);
		\draw[dashed] (0,0) circle (1);
		\node[above] at (0,1) {$\mathcal{D}_\epsilon$};
		\draw[<->,thick] (0,0)++(210:0.1) -- ++(210:0.8);
		\node at (-0.3,-0.4) {$\epsilon$};
		\draw[very thick,red,dashed] (6,0) node[black,right] {$\tau=\tau_0$} -- (1,0) node[pos=0.45,above right,black] {$A_{\tau_0}$};
		\draw[very thick,red] (1,0) arc (0:30:1) node[midway,right,black] {$B_\epsilon$} coordinate (C);
		\draw[very thick,red,dashed] (C) -- ++(30:5) coordinate (B) node[pos=0.55,below right,black] {$A_{\tau_1}$} node[black,right] {$\tau=\tau_1$};
		\fill[red] (0,0) circle (0.05);
		\node[above] (0,0) {$\Upsilon$};
%		\path (0,0) ++(15:4.2) node[scale=1.3] {$\mathcal{M}(\tau_0,\tau_1)$};
		\draw[->,thick] (5:6.3) arc (5:25:6.3);
	\end{tikzpicture}
\end{equation}
The action in this region can be written 
\begin{equation}
	S = \int_{\mathcal{M}(\tau_0,\tau_1)}*L + S_{\text{boundary}}.
\end{equation}
$S_{\text{boundary}}$ is a boundary term which only has relevant contributions away from the entangling surface, so we can ignore it in what follows. When we calculate the variation of the action and restrict to on-shell field configurations, the result is
\begin{align}
	\delta S &= \int_{\mathcal{M}(\tau_0,\tau_1)}\dd{\theta} + \delta S_{\text{boundary}} \\
	&= \int_{A_{\tau_1}}\theta - \int_{A_{\tau_0}}\theta - \int_{B_\epsilon} \theta + (\dots),
	\label{Equation: action variation symplectic potential contributions}
\end{align}
where the signs denote the orientations chosen, and the ellipsis in parentheses here and in the following contains terms away from $\Upsilon$ that we do not care about. Comparing to the classical mechanical case, we want to put this variation in the form $\delta S = \Theta(\tau_1) - \Theta(\tau_0)$, where $\Theta(\tau)$ is determined in terms of the fields at angle $\tau$. The tempting approach, and the one that is usually used in the covariant phase space method, is simply to set $\Theta(\tau) = \int_{A_\tau}\theta + (\dots)$, but of course this will not take account of the contribution at $B_\epsilon$, and will suffer from the $\theta\to\theta+\dd{Y}$ ambiguity noted previously.

We will supply a method to properly account for the contribution at $B_\epsilon$. Note that the expression \eqref{Equation: action variation symplectic potential contributions} is insensitive to $\theta\to\theta+\dd{Y}$. Therefore, once we have taken the contribution at $B_\epsilon$ into account, the ambiguity will have been dealt with.

Let $\Upsilon_{\epsilon,\tau} = \partial\mathcal{D}_\epsilon\cap\partial A_\tau$. These $(D-2)$-surfaces comprise a smooth foliation of $\partial\mathcal{D}_\epsilon=\bigcup_\tau B_{\epsilon,\tau}$. Each $\Upsilon_{\epsilon,\tau}$ can be viewed as a displacement of the entangling surface $\Upsilon$ by a distance $\epsilon$ in the direction of the angle $\tau$. We can decompose the contribution to $\delta S$ at $B_\epsilon$ into an integral over contributions at $\Upsilon_{\epsilon,\tau}$ as follows:
\begin{equation}
	\int_{B_\epsilon}\theta = \int_{\tau_0}^{\tau_1}\dd{\tau'}F(\tau') \qq{where} F(\tau') = \int_{\Upsilon_{\epsilon,\tau'}}\iota_{\partial_\tau}\theta.
\end{equation}

We assume that $\theta$ is smoothly defined near $\Upsilon$. Then we can expand $F(\tau')$ in a Fourier series in $\tau'$ that remains well-defined in the $\epsilon\to0$ limit. We write
\begin{equation}
	F(\tau') = \sum_{m=-\infty}^{\infty}f_me^{im\tau'},
\end{equation}
where
\begin{equation}
	f_m = \frac1{2\pi}\int_0^{2\pi}\dd{\tau'}F(\tau')e^{-im\tau'} = \frac1{2\pi}\int_{\partial \mathcal{D}_\epsilon}\theta e^{-im\tau}.
\end{equation}
Performing the $B_\epsilon$ integral, we thus have
\begin{equation}
	\int_{B_\epsilon}\theta = \qty[f_0 \tau'+\sum_{m\ne 0}\frac1{im}f_me^{im\tau'}]_{\tau'=\tau_0}^{\tau'=\tau_1}
\end{equation}
Substituting this into \eqref{Equation: action variation symplectic potential contributions}, we find that we can write $\delta S = \Theta(\tau_1) - \Theta(\tau_2)$, where
\begin{align}
	\Theta(\tau') &= C + \int_{A_{\tau'}}\theta - f_0\tau' - \sum_{m\ne 0}\frac1{im}f_me^{im\tau'} + \dots\\
	&= C + \int_{A_{\tau'}}\theta - \frac{\tau'}{2\pi}\int_{\partial \mathcal{D}_\epsilon}\theta - \frac1{2\pi}\sum_{m\ne 0}\frac1{im}e^{im\tau'}\int_{\partial \mathcal{D}_\epsilon}\theta e^{-im\tau} +(\dots).
	\label{Equation: unambiguified symplectic potential}
\end{align}
$C=C[\phi,\delta\phi_0]$ is the undetermined time-independent zero-mode term.
In the limit $\epsilon\to 0$, each of the integrals over $\partial\mathcal{D}_\epsilon$ become locally defined objects at the entangling surface. Therefore, \eqref{Equation: unambiguified symplectic potential} gives a good definition of the symplectic potential near $\Upsilon$ at the angle $\tau'$.

Armed with this definition, we can now obtain the symplectic structure by taking the field space exterior derivative of $\Theta$. The result is
\begin{equation}
	\Omega = U + \int_{A_{\tau'}}\omega - \frac{\tau'}{2\pi}\int_{\partial \mathcal{D}_\epsilon}\omega - \frac1{2\pi}\sum_{m\ne 0}\frac1{im}e^{im\tau'}\int_{\partial \mathcal{D}_\epsilon}\omega e^{-im\tau} + (\dots),
	\label{Equation: unambiguous symplectic form}
\end{equation}
where $\omega$ is defined in \eqref{Equation: symplectic form density}, and 
\begin{equation}
	U[\phi,\delta_1\phi_0,\delta_2\phi_0] = \delta_1C[\phi,\delta_2\phi_0] - \delta_2C[\phi,\delta_1\phi_0] - C[\phi,\delta_{12}\phi_0]
\end{equation}
is the undetermined zero-mode term that comes from taking the field space exterior derivative of $C$.\footnote{One can see that $U$ can only depend on the field variations through their zero modes by applying the same reasoning used to show that this was true for $C$.}

\subsection{Diffeomorphism charges}

We will assume in this section that the fields are on-shell. Consider an infinitesimal diffeomorphism parametrised by a vector field $\xi$. This diffeomorphism acts on the fields $\phi$ by Lie derivative, $\phi\to \phi + \lie_\xi\phi$. If we can find a function $H_\xi$ on phase space such that
\begin{equation}
	\delta H_\xi[\phi]=\Omega[\phi,\delta\phi,\lie_\xi\phi],
	\label{Equation: generator definition}
\end{equation}
then $H_\xi$ is the Hamiltonian charge which generates the diffeomorphism parametrised by $\xi$. In this subsection we will evaluate the right-hand side of \eqref{Equation: generator definition}.

Substituting $\delta \phi = \lie_\xi\phi$ into the on-shell relation $\delta(*L) = \dd{\theta}$, we find
\begin{equation}
	\dd{\big(\iota_\xi(*L(\phi))\big)} = \dd(\theta(\phi,\lie_\xi\phi)).
\end{equation}
Therefore, $\theta(\phi,\lie_\xi\phi) - \iota_\xi(*L(\phi))$ is closed for all $\xi$. Furthermore, it vanishes for $\xi=0$, and so by the results of~\cite{Wald:2010:doi:10.1063/1.528839} it must be exact. Hence we can write
\begin{equation}
	\theta(\phi,\lie_\xi\phi) - \iota_\xi(*L(\phi)) = \dd(Q_\xi(\phi))
\end{equation}
for some $(D-2)$-form $Q_\xi$, which is known as the Noether charge density. $Q_\xi$ is defined up to the addition of an exact form, $Q_\xi \to Q_\xi + \dd{Z}$. 

If we write $\delta_1\phi=\delta\phi$ and $\delta_2\phi = \lie_\xi\phi$, then we have
\begin{equation}
	\delta_{12}\phi=[\delta_1,\delta_2]\phi = \delta(\lie_\xi\phi) - \lie_\xi(\delta\phi) = \lie_{\delta \xi}\phi
	\label{Equation: commutated diffeomorphism}
\end{equation}
So the variation $\delta_{12}\phi$ is equivalent to an infinitesimal diffeomorphism parametrised by the vector field $\delta \xi$. Note that in general $\xi$ is allowed to depend on $\phi$, so it is possible to have $\delta \xi\ne 0$. 

We can use \eqref{Equation: commutated diffeomorphism} to obtain
\begin{align}
	\omega(\phi,\delta\phi,\lie_\xi\phi) &= \delta\theta(\phi,\lie_\xi\phi) - \lie_\xi\theta(\phi,\delta\phi) - \theta(\phi,\lie_{\delta \xi}\phi)\\
	\begin{split}
		&=\delta\big(\iota_\xi(*L(\phi)) + \dd(Q_\xi(\phi))\big) 
		- \iota_\xi\dd{\theta(\phi,\delta\phi)} - \dd{\big(\iota_\xi\theta(\phi,\delta\phi)\big)}\\
		&\qquad\qquad- \iota_{\delta\xi}(*L(\phi)) - \dd(Q_{\delta\xi}(\phi))
	\end{split}\\
	&= \dd{\big(\delta Q_\xi(\phi) - Q_{\delta\xi}(\phi) -\iota_\xi\theta(\phi,\delta\phi)\big)} +\iota_\xi\big(\underbrace{\delta(*L)-\dd(\theta(\phi,\delta\phi))}_{=0}\big).
\end{align}
So $\omega(\phi,\delta\phi,\lie_\xi\phi)$ is equal to the exact form given on the last line above. Substituting this into \eqref{Equation: unambiguous symplectic form}, we find
\begin{multline}
	\Omega[\phi,\delta\phi,\lie_\xi\phi] = U[\phi,\delta\phi_0,\lie_\xi\phi_0] + \int_{\Upsilon_{\epsilon,\tau'}} \delta Q_\xi(\phi) - Q_{\delta\xi}(\phi) -\iota_\xi\theta(\phi,\delta\phi) \\
	- \frac1{2\pi}\sum_{m\ne 0}\frac1{im}e^{im\tau'}\int_{\partial \mathcal{D}_\epsilon}\dd{\big(\delta Q_\xi(\phi) - Q_{\delta\xi}(\phi) -\iota_\xi\theta(\phi,\delta\phi)\big)} e^{-im\tau} + (\dots).
\end{multline}
At this stage we can note that this equation is independent of the ambiguity $Q_\xi\to Q_\xi+\dd{Z}$.
A partial integration on the second line gives
\begin{multline}
	\Omega[\phi,\delta\phi,\lie_\xi\phi] = U[\phi,\delta\phi_0,\lie_\xi\phi_0] + \int_{\Upsilon_{\epsilon,\tau'}} \delta Q_\xi(\phi) - Q_{\delta\xi}(\phi) -\iota_\xi\theta(\phi,\delta\phi) \\
	- \frac1{2\pi}\sum_{m\ne 0}e^{im\tau'}\int_{\partial \mathcal{D}_\epsilon}\dd{\tau}\wedge\big(\delta Q_\xi(\phi) - Q_{\delta\xi}(\phi) -\iota_\xi\theta(\phi,\delta\phi)\big) e^{-im\tau} + (\dots).
	\label{Equation: charge variation almost}
\end{multline}
We can expand the latter term on the first line in a Fourier series as
\begin{equation}
	\int_{\Upsilon_{\epsilon,\tau'}} \delta Q_\xi(\phi) - Q_{\delta\xi}(\phi) -\iota_\xi\theta(\phi,\delta\phi) = \sum_{m=-\infty}^{\infty} e^{im\tau'} h_m,
\end{equation}
where
\begin{align}
	h_m &= \frac1{2\pi}\int_0^{2\pi}\dd{\tau'}e^{-im\tau'}\int_{\Upsilon_{\epsilon,\tau'}}\delta Q_\xi(\phi) - Q_{\delta\xi}(\phi) -\iota_\xi\theta(\phi,\delta\phi)\\
	&=\frac1{2\pi}\int_0^{2\pi}\dd{\tau'}e^{-im\tau'}\int_{\Upsilon_{\epsilon,\tau'}}\iota_{\partial_\tau}\big[\dd{\tau}\wedge\big(\delta Q_\xi(\phi) - Q_{\delta\xi}(\phi) -\iota_\xi\theta(\phi,\delta\phi)\big)\big]\\
	&=\frac1{2\pi}\int_{\partial \mathcal{D}_\epsilon}\dd{\tau}\wedge\big(\delta Q_\xi(\phi) - Q_{\delta\xi}(\phi) -\iota_\xi\theta(\phi,\delta\phi)\big) e^{-im\tau}.
\end{align}
Noting that $-h_m$ is exactly the term that is summed over in the second line of \eqref{Equation: charge variation almost}, we see that everything cancels except for the summand at $m=0$. Therefore, all that remains of the Fourier series expansion is $h_0$. We thus have
\begin{equation}
	\Omega[\phi,\delta\phi,\lie_\xi\phi] = U[\phi,\delta\phi_0,\lie_\xi\phi_0] + \frac1{2\pi}\int_{\partial \mathcal{D}_\epsilon}\dd{\tau}\wedge\big(\delta Q_\xi(\phi) - Q_{\delta\xi}(\phi) -\iota_\xi\theta(\phi,\delta\phi)\big).
	\label{Equation: charge variation}
\end{equation}

\section{The entangling surface boost generator}
\label{Section: Entangling surface charge}

The particular Hamiltonian charge we are interested in is $iK$, which generates Euclidean rotations about the entangling surface $\Upsilon$. This can be obtained by setting $\xi=\partial_\tau=i\partial_\eta$ in \eqref{Equation: charge variation}. This particular $\xi$ is independent of the fields $\phi$, so we have $\delta\xi=0$. Also, note that $\lie_{\partial_\tau}\phi_0 = 0$; this is after all the definition of a zero mode. Therefore, we can set $U[\phi,\delta\phi_0,\lie_{\partial_\tau}\phi_0]=0$. Thus, from here on, the ambiguous term $U$ in the symplectic structure will not have any impact on our calculations.

We have
\begin{equation}
	i\delta K[\phi] = \delta H_{\partial_\tau} = \frac1{2\pi}\int_{\partial \mathcal{D}_\epsilon}\dd{\tau}\wedge\big(\delta Q_{\partial_\tau}(\phi) -\iota_{\partial_\tau}\theta(\phi,\delta\phi)\big) + (\dots).
\end{equation}
Using $\dd{\tau}\wedge(\iota_{\partial_\tau\theta}) = \theta-\iota_{\partial_\tau}(\dd{\tau}\wedge\theta)$, and the fact that $\partial_\tau$ is tangential to $\partial \mathcal{D}_\epsilon$, this can be written
\begin{align}
	i\delta K[\phi] &= \frac1{2\pi}\int_{\partial \mathcal{D}_\epsilon}\delta(\dd{\tau}\wedge Q_{\partial_\tau}(\phi)) + \theta(\phi,\delta\phi) + (\dots)\\
	&= \frac1{2\pi}\int_{\partial \mathcal{D}_\epsilon}\delta(\dd{\tau}\wedge Q_{\partial_\tau}(\phi)) + \frac1{2\pi}\int_{\mathcal{D}_\epsilon} \dd(\theta(\phi,\delta\phi)) + (\dots)\\
	&= \delta\qty(\frac1{2\pi}\int_{\partial \mathcal{D}_\epsilon}\dd{\tau}\wedge Q_{\partial_\tau}(\phi) + \frac1{2\pi}\int_{\mathcal{D}_\epsilon}*L(\phi)) + (\dots).
\end{align}
In the last line we used $\delta(*L)=\dd{\theta}$. Therefore, $K$ can be written, up to an irrelevant constant, as
\begin{equation}
	iK = i\tilde{K} + \frac1{2\pi}\int_{\partial \mathcal{D}_\epsilon}\dd{\tau}\wedge Q_{\partial_\tau} + \frac1{2\pi}\int_{\mathcal{D}_\epsilon}*L,
	\label{Equation: boot charge almost}
\end{equation}
where $\tilde{K}$ contains contributions that do not originate at the entangling surface.
Since the Lagrangian density is assumed to be smooth at $\Upsilon$, in the limit $\epsilon\to 0$ the term $\frac1{2\pi}\int_{\mathcal{D}_\epsilon}*L\to 0$, so we will ignore it in the following.

Using similar manipulations to previously, this can now be put into the form
\begin{equation}
	K_\Upsilon = \lim_{\epsilon \to 0}\frac1{2\pi}\int_0^{2\pi}\dd{\tau'}\int_{\Upsilon_{\epsilon,\tau'}}Q_{\partial_\eta},
	\label{Equation: angle average boost charge}
\end{equation}
where we have defined $K_\Upsilon=K-\tilde{K}$, substituted in $\partial_\tau=i\partial_\eta$, cancelled the factor of $i$, and now choose to explicitly include the limit $\epsilon\to 0$.

We perhaps should have expected the apparent averaging over $\tau'$ in the above expression. After all, the density matrix near $\Upsilon$ is thermal, and it is an elementary result in equilibrium statistical mechanics that ensemble expectation values are equivalent to time averaged expectation values.

It is desirable to have an expression for $K_\Upsilon$ completely in terms of the boost parameter $\eta$ instead of the angle $\tau$. We can achieve this by analytically continuing $\int_{\Upsilon_{\epsilon,\tau'}}Q_{\partial_\eta}$ to complex $\tau'$. If we let $z=e^{i\tau'}$, then we can write the above expression as a contour integral in the complex $z$-plane. We have
\begin{equation}
	K_\Upsilon = \lim_{\epsilon\to0}\frac1{2\pi i}\oint_\gamma\frac{\dd{z}}{z}\int_{\Upsilon_{\epsilon,\tau'}}Q_{\partial_\eta},
	\label{Equation: boost contour integral}
\end{equation}
where $\gamma$ is the contour that goes once around $\abs{z}=1$. We will assume that $\int_{\Upsilon_{\epsilon,\tau'}}Q_{\partial_\eta}$ is free of singularities in the interior of $\gamma$.\footnote{We feel this that this a sensible assumption to make at this point in the analysis, but will briefly comment on how it could be violated. The presence of singularities inside $\gamma$ would indicate non-smooth Lorentzian evolution. This would arise from the transit of topologically non-trivial excitations across the entangling surface. Such excitations are generally charged under the action of the boost. The resulting contributions to the contour integral from the associated poles would thus account for the charges of these excitations. We leave the full exploration of this to future work.} Then the single contribution to the contour integral comes from $z=0$, which is reached by sending $\eta'=i\tau'\to-\infty$, and we can write
\begin{equation}
	K_\Upsilon = \lim_{\eta'\to-\infty,\epsilon\to 0} \int_{\Upsilon_{\epsilon,-i\eta'}}Q_{\partial_\eta}.
	\label{Equation: infinite boost boost charge}
\end{equation}
$\Upsilon_{\epsilon,-i\eta}$ is a surface which has been Lorentz boosted by an amount $\eta$. Thus $K_\Upsilon$ can be evaluated by calculating the integral of $Q_{\partial_\eta}$ over an infinitely boosted version of the entangling surface. 

The order of the two limits $\eta'\to -\infty$, $\epsilon \to 0$ is important. The implications of different orderings are easiest to understand by visualising the action of a boost in Lorentzian spacetime. This is portrayed below.
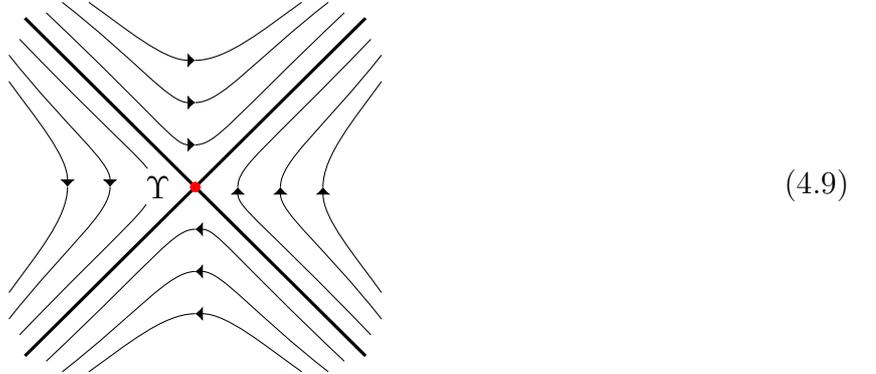
\begin{equation}
	\begin{tikzpicture}[baseline={([yshift=-.5ex]current bounding box.center)},scale=0.7]
        \draw[very thick] (-3.2,-3.2) -- (3.2,3.2);
        \draw[very thick] (3.2,-3.2) -- (-3.2,3.2);
        \fill[red] (0,0) circle (0.1);

        \newcommand{\boostflow}{
            \draw (3.3,2.8) .. controls (1.3,0.7) and (0.8,0.4) .. (0.8,0);
            \draw (3.5,2.5) .. controls (2.4,1.1) and (1.6,0.5) .. (1.6,0);
            \draw (3.5,2) .. controls (3,1.3) and (2.4,0.6) .. (2.4,0);
        }
        \foreach \x in {0,1,2,3}
        {
            \begin{scope}[rotate={90*\x},yscale={-(-1)^\x},-{Latex[length=1mm,width=2mm]}]
                \boostflow
            \end{scope}
            \begin{scope}[rotate={90*\x},yscale={(-1)^\x}]
                \boostflow
            \end{scope}
        }

        \fill[white] (-0.7,0) circle (0.4);
        \node at (-0.7,0) {$\Upsilon$};
    \end{tikzpicture}
\end{equation}
The two diagonal lines represent the two sets of null rays normal to $\Upsilon$, which is the surface at which they intersect. The action of a boost is shown by the curved lines. 

Suppose we were to take $\epsilon\to 0$ first. Then $\Upsilon_{\epsilon,-i\eta'}$ would coincide with $\Upsilon$. The action of the boost is vanishing at $\Upsilon$. Therefore, after carrying out the limit $\eta'\to-\infty$, $\Upsilon_{\epsilon,-i\eta'}$ would still coincide with $\Upsilon$. On the other hand, suppose we keep $\epsilon$ small but non-zero, and start by taking the limit $\eta'\to-\infty$. Then the surface $\Upsilon_{\epsilon,-i\eta'}$ would flow along the action of the boost, which is non-trivial for $\epsilon \ne 0$. After infinitely boosting, and subsequently taking $\epsilon \to 0$, the surface $\Upsilon_{\epsilon,-i\eta'}$ would end up infinitely propagated along one of the sets of null rays normal to $\Upsilon$. These two scenarios are depicted below.
\begin{equation}
    \begin{array}{ccc}
        \begin{tikzpicture}[baseline={([yshift=-.5ex]current bounding box.center)},scale=0.7]
            \draw[very thick] (-3.2,-3.2) -- (3.2,3.2);
            \draw[very thick] (3.2,-3.2) -- (-3.2,3.2);
            \fill[red] (0,0) circle (0.1);
            \node[left] at (-0.3,0) {$\Upsilon_{0,-i\eta'} = \Upsilon_{0,i\infty} = \Upsilon$};

            \draw[thick,-{Latex[length=1mm,width=2mm]}] (1.6,0) -- (0.2,0);
            \fill[red] (1.6,0) circle (0.1) node[right,black] {$\Upsilon_{\epsilon,-i\eta'}$};
        \end{tikzpicture}
        &\qquad\qquad&
        \begin{tikzpicture}[baseline={([yshift=-.5ex]current bounding box.center)},scale=0.7]
            \draw[very thick] (-3.2,-3.2) -- (3.2,3.2);
            \draw[very thick] (3.2,-3.2) -- (-3.2,3.2);
            \fill[red] (0,0) circle (0.1);
            \node at (-0.7,0) {$\Upsilon$};

            \draw[thick,{Latex[length=1mm,width=2mm]}-] (3.7,2.7) .. controls (2.4,1.1) and (1.6,0.5) .. (1.6,0);

            \fill[red] (1.6,0) circle (0.1) node[right,black] {$\Upsilon_{\epsilon,-i\eta'}$};
            \draw[thick,-{Latex[length=1mm,width=2mm]}] (3.8,2.8) -- (3.4,3.2);

            \fill[red] (3.8,2.8) circle (0.1) node[right,black] {$\Upsilon_{\epsilon,i\infty}$};
            \fill[red] (3.3,3.3) circle (0.1) node[above,black] {$\Upsilon_{0,i\infty}$};
        \end{tikzpicture}
        \\
        \\
        \text{(a) $\epsilon\to0$ then $\eta'\to-\infty$.} && \text{(b) $\eta'\to-\infty$ then $\epsilon\to0$.}
    \end{array}
    \label{Diagram: non-commutativity of limits}
\end{equation}
It should be clear that, by carefully tuning the relative speeds of the two limits, we can have $\Upsilon_{\epsilon,-i\eta'}$ end up at different points along one of the four sets of null normal rays originating at $\Upsilon$.

Because of its simplicity, it is tempting to choose option (a) in \eqref{Diagram: non-commutativity of limits}. However, the way in which we are computing the boost charge seems to imply that we need to choose option (b), since the $\epsilon \to 0$ limit ought to be taken \emph{after} doing the contour integral. Unfortunately, this may lead to divergences, due to the infinite null propagation, and it will be necessary to find an appropriate regularisation of these divergences. We have not yet found an appropriate regularisation procedure, and will leave this to future work. 

It is worth briefly mentioning that in the case where the spacetime fields are boost-invariant, this ordering ambiguity will not have an impact. This is because the limit $\eta'\to-\infty$ is trivial, since all fields are independent of $\eta'$. In the boost-dependent case, we expect that the formulae \eqref{Equation: angle average boost charge} and \eqref{Equation: infinite boost boost charge} should pick out some kind of boost-averaged version of $\int Q_{\partial_\eta}$.

In the following two subsections we will evaluate $K_\Upsilon$ first in general relativity, and then in higher derivative gravity theories.

\subsection{General relativity}
\label{Subsection: GR boost charge}

Pure general relativity is described by the Einstein-Hilbert action \eqref{Equation: Einstein-Hilbert}. The corresponding Lagrangian density is $*L = \frac1{16\pi G} {*R}$, and it can be shown that the Noether charge density can be chosen to take the form $Q_\xi = \frac1{16\pi G}{*\dd(g(\xi))}$, where $g(\xi)$ is the 1-form obtained by application of the metric to $\xi$.

We can  write the Euclidean line element in the effective spacetime near $\Upsilon$ as
\begin{equation}
	\dd{s}^2|_{\Upsilon} = \dd{r}^2 + r^2\dd{\tau}^2 + q_{AB}\dd{\sigma^A}\dd{\sigma^B}.
\end{equation}
Here $r,\tau$ are radial coordinates in the normal plane to $\Upsilon$, the $\sigma^A$, $A=2,\dots,D-1$ are a set of coordinates on the level surfaces of constant $r,\tau$, and $q_{AB}$ are the components in these coordinates of the induced metric on these level surfaces. In these coordinates, $\mathcal{D}_\epsilon$ is defined as the region $r \le \epsilon$, and $\Upsilon_{\epsilon,\tau'}$ is the level surface at $r=\epsilon,\tau=\tau'$.

Analytically continuing to imaginary $\tau=i\eta$, the Lorentzian line element takes the form
\begin{equation}
	\dd{s}^2|_{\Upsilon} = \dd{r}^2 - r^2\dd{\eta}^2 + q_{AB}\dd{\sigma^A}\dd{\sigma^B}.
	\label{Equation: line element near entangling surface}
\end{equation}
Setting $\xi=\partial_\eta$, one readily finds that $g(\xi)=-r^2\dd{\eta}$ which implies that $\dd(g(\xi))=2r\dd{\eta}\wedge\dd{r}$. Application of the Hodge star then gives $\hodge\dd(g(\xi)) = 2\sqrt{\det q} \dd{\sigma^2}\wedge\dots\wedge\dd{\sigma^{D-1}}$. 

Using this in \eqref{Equation: angle average boost charge}, we may write
\begin{align}
	K_\Upsilon &= \frac1{2\pi}\int_0^{2\pi}\dd{\tau} \lim_{\epsilon\to 0}\frac1{8\pi G}\int_{\Upsilon_{\epsilon,\tau'}}\dd[D-2]{\sigma}\sqrt{\det q}\\
	&= \frac1{2\pi}\int_0^{2\pi}\dd{\tau'} \frac1{8\pi G}\underbrace{\lim_{\epsilon\to 0} \mathcal{A}[\Upsilon_{\epsilon,\tau'}]}_{=\mathcal{A}[\Upsilon]}.
\label{Equation: entangling surface boost charge}
\end{align}
In this expression, $\mathcal{A}[X]$ denotes the area of $X$.

We therefore have found that in general relativity the boost charge at the entangling surface is given by $K_\Upsilon = \frac1{8\pi G}\mathcal{A}[\Upsilon]$.\footnote{This quantity may be divergent, in which case it will need to be regularised. We will not carry out such a regularisation here.} We used above the fact that in the limit $\epsilon\to 0$, the area of $\Upsilon_{\epsilon,\tau'}$ loses any dependence on $\tau'$, and converges to $\mathcal{A}[\Upsilon]$. This independence of $\tau'$ is only a consequence of the particular theory of gravity we are considering. One should not expect $\int_{\Upsilon_{\epsilon,\tau'}}Q_{\partial_\eta}$ to be independent of $\tau'$ in general.

\subsection{Higher derivative gravity}
\label{Subsection: higher derivative boost charge}

Now suppose the Lagrangian density is constructed locally from the metric $g_{ab}$, the Riemann tensor $R_{abcd}$, and arbitrarily many symmetrised covariant derivatives of the Riemann tensor. In~\cite{Iyer:1994ys} it was shown that for any local theory of gravity without additional matter fields,\footnote{For simplicity we shall only consider gravity in the absence of additional matter fields, but the extension to include these fields should be straightforward.} $L$ can be put in this form, and that the Noether charge density for such a theory may be written
\begin{equation}
	Q_\xi = \iota_\xi W - \hodge(\dd{x^a}\wedge\dd{x^b})E\indices{_{ab}^{cd}}\nabla_{[c}\xi_{d]},
	\label{Equation: higher derivative Noether}
\end{equation}
where $W$ is some local geometry-dependent $(D-1)$-form, and
\begin{align}
	E^{abcd} = \fdv{L}{R_{abcd}} &= \pdv{L}{R_{abcd}} - \nabla_e\pdv{L}{(\nabla_eR_{abcd})} + \nabla_{(e}\nabla_{f)}\pdv{L}{(\nabla_{(e}\nabla_{f)}R_{abcd})} - \dots \\
	&= \sum_m (-1)^m\nabla_{(e_1}\dots\nabla_{e_m)}\pdv{L}{(\nabla_{(e_1}\dots\nabla_{e_m)}R_{abcd})}
\end{align}
is the Euler-Lagrange derivative of $L$ with respect to the Riemann tensor. The partial derivatives in this expression are evaluated by treating the Riemann tensor and its derivatives as independent of each other and the metric, and are uniquely defined so that they have the same tensor symmetries as the varied quantities.

Using \eqref{Equation: higher derivative Noether} in \eqref{Equation: angle average boost charge}, we find
\begin{align}
	K_\Upsilon &= \lim_{\epsilon\to 0}\frac1{2\pi}\int_{\partial \mathcal{D}_\epsilon}\dd{\tau}\wedge\big(\iota_\xi W - \hodge(\dd{x^a}\wedge\dd{x^b})E\indices{_{ab}^{cd}}\nabla_{[c}\xi_{d]}\big)\\
	&= - \lim_{\epsilon\to 0}\frac1{2\pi} \int_0^{2\pi}\dd{\tau'}\int_{\Upsilon_{\epsilon,\tau'}} \hodge(\dd{x^a}\wedge\dd{x^b})E\indices{_{ab}^{cd}}\nabla_{[c}\xi_{d]} - \lim_{\epsilon\to 0}\frac1{2\pi}\int_{\mathcal{D}_\epsilon} W,
\end{align}
where $\xi = \partial_\eta$. $W$ is smooth in $\mathcal{D}_\epsilon$, so in the limit $\epsilon\to0$, we may discard the integral $\int_{\mathcal{D}_\epsilon} W$ (this is the same reasoning that was used to discard $\int_{\mathcal{D}_\epsilon}*L$ in \eqref{Equation: boot charge almost}).

The pullback of $*(\dd{x^a}\wedge\dd{x^b})$ to $\Upsilon_{\epsilon,\tau'}$ is given by
\begin{equation}
	*(\dd{x^a}\wedge\dd{x^b})|_{\Upsilon_{\epsilon,\tau'}} = \frac12\epsilon^{ab}\sqrt{\det q} \dd{\sigma^2}\wedge\dots\wedge\dd{\sigma^{D-1}},
\end{equation}
where $\epsilon^{ab} = \frac1{2r}(\delta^a_\eta\delta^b_r - \delta^a_r\delta^b_\eta)$. Also, we have
\begin{equation}
	\nabla_{[c}\xi_{d]}\dd{x^c}\wedge\dd{x^d} = \dd(g(\xi)) = 2r\dd{\eta}\wedge\dd{r} \implies \nabla_{[c}\xi_{d]} = 2\epsilon_{cd}.
\end{equation}
Thus, putting things in the form \eqref{Equation: infinite boost boost charge}, the boost charge at the entangling surface can be written
\begin{equation}
	K_\Upsilon = \lim_{\eta'\to-\infty,\epsilon\to 0}\frac{\hbar}{2\pi} S_{\text{Wald}}[\Upsilon_{\epsilon,-i\eta'}],
    \label{Equation: higher derivative boost charge pre limit}
\end{equation}
where
\begin{equation}
	S_{\text{Wald}}[\Upsilon_{\epsilon,-i\eta'}] = -\frac{2\pi}\hbar\int_{\Upsilon_{\epsilon,-i\eta'}}\dd[D-2]{\sigma}\sqrt{\det q}\fdv{L}{R_{abcd}}\epsilon_{ab}\epsilon_{cd}
\end{equation}
is equal to the Wald entropy functional \eqref{Equation: Wald entropy}~\cite{Wald:1993nt} evaluated on the surface $\Upsilon_{\epsilon,-i\eta'}$.

At this point, the limit ordering ambiguity previously mentioned becomes important. Choosing (a) in \eqref{Diagram: non-commutativity of limits} would lead to $K_\Upsilon = \frac\hbar{2\pi} S_{\text{Wald}}[\Upsilon]$. However, as previously discussed, we ought to instead choose something closer to (b). In that case, with an appropriate regularisation procedure in hand, the answer we get should be something like the Wald entropy associated to the boost-invariant part of the fields. The correct method for calculating the boost-invariant part of the fields depends upon the exact regularisation procedure used. The resulting quantity is clearly related to the Iyer-Wald dynamical entropy $S_{\text{Iyer-Wald}}[\Upsilon]$~\cite{Iyer:1994ys} of the entangling surface, but there is no guarantee that the two are equal.

We shall assume in this paper that we have chosen a regularisation procedure, and can therefore take the limit in \eqref{Equation: higher derivative boost charge pre limit} in a well-defined way. We can then write
\begin{equation}
	K_\Upsilon = \frac{\hbar}{2\pi}S_{\text{dyn}}[\Upsilon],
	\label{Equation: higher derivative boost charge}
\end{equation}
where
\begin{equation}
    S_{\text{dyn}}[\Upsilon] = \lim_{\eta'\to-\infty,\epsilon\to 0} S_{\text{Wald}}[\Upsilon_{\epsilon,-i\eta'}]
\end{equation}
is a dynamical version of the Wald entropy.

\section{Minimal surfaces and the semiclassical limit}
\label{Section: Semiclassical limit}

The semiclassical limit is defined as $\hbar\to 0$. Recall that the elements of the reduced density matrix are given by \eqref{Equation: reduced density decomposition}, which is repeated below for convenience:
\begin{equation*}
	\mel*{\gamma^1_A}{\rho_A}{\gamma^0_A} = \mel*{\gamma^1_A}{e^{-\frac1\hbar\pi K} \mathscr{U} e^{-\frac1\hbar\pi K}}{\gamma^0_A}.
\end{equation*}
Ignoring $\mathscr{U}$, in the limit $\hbar\to 0$, these matrix elements are clearly dominated by states which minimise $K$. If we further ignore $\tilde{K}$, i.e.\ the contributions to $K$ which do not originate at the entangling surface, then we find that the matrix elements are dominated by those states which minimise the entangling surface boost charge $K_\Upsilon$. Combining this with the results of the previous section, we conclude that in general relativity the matrix elements are dominated by those states for which the area of the entangling surface is minimised, and in higher derivative gravity they are dominated by those states for which the dynamical entropy of the entangling surface is minimised.

The reader may be concerned about the validity of the choices just made to ignore contributions away from $\Upsilon$. We will now describe a way in which this validity can be controlled.

Consider again the original, unreduced density matrix $\rho$. It is a fundamental requirement in all theories of gravity that the states defined on a surface $\Sigma$ be invariant under `small' diffeomorphisms, i.e.\ diffeomorphisms with trivial action at $\partial\Sigma$ (conversely a `large' diffeomorphism is one whose action is non-trivial at $\partial\Sigma$). The action is certainly invariant under small diffeomorphisms. Therefore, in order to guarantee that the path integral in \eqref{Equation: density matrix elements} is similarly invariant, the measure $\Dd{g}$ must give equal weight to two metrics $g_1,g_2$, if those two metrics are related by a small diffeomorphism.

This means that we can factorise the measure into two components $\Dd{g} = \Dd{[g]}\Dd{\alpha}$. Each $[g]$ is an equivalence class of metrics modulo small diffeomorphisms, and each $\alpha$ is a small diffeomorphism. $\alpha$ then determines the metric $g$ as a certain representative of $[g]$.

In the path integral, the small diffeomorphism invariance means we can factor out the $\Dd{\alpha}$ integral. This then just contributes a constant factor in front of the path integral which cancels when we compute expectation values.

However, this is no longer the case after we have carried out the reduction procedure. In particular, there will exist diffeomorphisms $\alpha$ which were small in the original system, but which have non-trivial action at the entangling surface. Since, after reduction, the entangling surface is one part of the boundary of the surface on which states are measured, such diffeomorphisms must be considered to be large in the reduced system.

Consider the group $G_\Upsilon$ of all such small diffeomorphisms made large. The only part of the boundary where such diffeomorphisms have a non-trivial action is at the entangling surface. So far we have been using an `active' viewpoint, in which the diffeomorphisms are understood as acting on the fields. It is now useful to switch to a `passive' viewpoint, in which the diffeomorphisms do not change the fields but instead deform the entangling surface $\Upsilon$. The two viewpoints are physically equivalent.

Since the original path integral included an integration over all small diffeomorphisms $\alpha$, the reduced path integral must include an integration over the action of the group $G_\Upsilon$ on $\Upsilon$. In other words, deformations of $\Upsilon$ are a genuine degree of freedom in the reduced path integral. Furthermore, this degree of freedom is decoupled from other degrees of freedom.

We should clarify exactly which deformations of $\Upsilon$ are included in the group $G_\Upsilon$. Deformations of $\Upsilon$ which do not preserve $\partial\Upsilon$ would not correspond to small diffeomorphisms in the original unreduced path integral, so these are not allowed. It is also natural to restrict to deformations that are continuously connected to the identity, because our derivation of the boost charge only really holds within a connected component of phase space. This restriction means that the deformations in $G_\Upsilon$ can only move $\Upsilon$ about within a particular homology class. All deformations of $\Upsilon$ in $G_\Upsilon$ which obey these constraints are permitted and are therefore integrated over in the reduced path integral. 

So return again to the issue of dominant contributions in the semiclassical limit. The action of $G_\Upsilon$ is trivial away from $\Upsilon$. This means that, if we only care about dominant contributions with respect to the action of $G_\Upsilon$, then it is valid to ignore $\mathscr{U}$ and $\tilde{K}$. Therefore, the more precise statement of what happens in the semiclassical limit is the following: the matrix elements of the reduced density matrix are dominated by those for which $\Upsilon$ has been deformed by some element in $G_\Upsilon$ so that $K_\Upsilon$ is minimal.

To close this section, we define the minimum boost charge operator 
\begin{equation}
	K_{\Upsilon,\text{min}} = \min_{\alpha \in G_\Upsilon} K_{\alpha(\Upsilon)}.
\end{equation}
Here $K_{\alpha(\Upsilon)}$ just denotes what the boost charge of the entangling surface would be if it was deformed by $\alpha$. For the reasons discussed above, in the semiclassical limit we can at leading order in $\hbar$ replace $K_\Upsilon \to K_{\Upsilon,\text{min}}$.

\section{Computing the entropy}
\label{Section: entropy}

Now we will calculate the entropy associated to the subregion $A$. The appropriate definition of entropy is the von Neumann entropy of the reduced density matrix $\rho_A$, which is defined as
\begin{equation}
	S_A = -\tr(\hat\rho_A\log \hat\rho_A) = -\int \Dd{\gamma_A} \mel{\gamma_A}{\hat\rho_A\log\hat\rho_A}{\gamma_A}.
\end{equation}
In this expression $\hat\rho_A=\frac{\rho_A}{\mathcal{Z}_A}$ is the normalised density matrix, where $\mathcal{Z}_A$ is the partition function for the reduced density matrix. Trivially, $\mathcal{Z}_A$ is equal to the partition function $\mathcal{Z}$ for the original density matrix.
Let $e^{-\mathscr{W}} = e^{-\frac1\hbar\pi\tilde{K}}\mathscr{U} e^{-\frac1\hbar\pi\tilde{K}}$. Noting that 
\begin{equation}
	-\mel{\gamma_A}{\hat\rho_A\log\hat\rho_A}{\gamma_A} = \mel{\gamma_A}{\hat\rho_A\qty(\frac{2\pi K_\Upsilon}{\hbar} + \mathscr{W}+\log\mathcal{Z})}{\gamma_A},
\end{equation}
and using the definition of the expectation value $\expval{\mathcal{O}}_A = \tr(\hat\rho_A\mathcal{O})$ of an operator $\mathcal{O}$, we can write the entropy as
\begin{equation}
	S_A = \frac{2\pi \expval{K_\Upsilon}_A}{\hbar} + \expval{\mathscr{W}}_A + \log\mathcal{Z}.
\end{equation}
Consider the entropy $S$ of the non-reduced density matrix $\rho$. By a similar calculation to the above, we have
\begin{equation}
	S = -\tr(\hat\rho\log\hat\rho) = -\expval{\log\rho} + \log\mathcal{Z},
\end{equation}
where $\hat\rho=\frac{\rho}{\mathcal{Z}}$, and $\expval{\mathcal{O}}=\tr(\hat\rho\mathcal{O})$. Thus we can write
\begin{equation}
	S_A - S = \underbrace{\frac{2\pi \expval{K_\Upsilon}_A}{\hbar}}_{S_\Upsilon} + \big(\expval{\mathscr{W}}_A + \expval{\log\rho}\big).
\end{equation}
So the reduction procedure has increased the entropy by the amount on the right-hand side. Since we are discarding information about the degrees of freedom in the region $\bar{A}$, such an increase is to be expected. The first term $S_\Upsilon$ represents the contribution of entanglement across $\Upsilon$. The term in brackets represents the contribution of a loss of knowledge about bulk degrees of freedom in $\bar{A}$. It can be understood as a consequence of the conversion of the internal energy in $\bar{A}$ into heat.

As discussed in Section \ref{Section: Semiclassical limit}, at leading order in the semiclassical limit we can replace $K_\Upsilon\to K_{\Upsilon,\text{min}}$. Thus, at leading order, we have
\begin{equation}
	S_\Upsilon = \frac{2\pi \expval{K_{\Upsilon,\text{min}}}_A}{\hbar},
\end{equation}
That is, the contribution to the entropy of the region $A$ arising from entanglement across the surface $\Upsilon$ is given to leading order in $\hbar$ by the expectation value of $\frac{2\pi}{\hbar}K_{\Upsilon,\text{min}}$.

\subsection{Consistency with the Ryu-Takayanagi conjecture}

In the case of general relativity, we showed in Section \ref{Subsection: GR boost charge} that $K_\Upsilon = \frac{\mathcal{A}[\Upsilon]}{8\pi G}$. Thus, the contribution to the entropy from entanglement across $\Upsilon$ is proportional at leading order to the expectation value of the minimal area. Precisely:
\begin{equation}
	S_\Upsilon = \frac{\expval{\mathcal{A}_{\text{min}}}_A}{4 G\hbar} = \frac1{4G\hbar}\expval{\min_{\alpha\in G_\Upsilon}\mathcal{A}[\alpha(\Upsilon)]}_A,
\end{equation}
where $G_\Upsilon$ is the group of entangling surface deformations described in Section \ref{Section: Semiclassical limit}.

This is exactly the value conjectured by Ryu and Takayanagi. It is interesting to note that although the Ryu-Takayanagi conjecture was originally supposed to only be relevant for AdS/CFT, the result obtained here holds for \emph{any} spacetimes, regardless of their boundary conditions, and is independent of the holographic principle.

\subsection{Higher derivative gravity}

In the case of higher derivative gravity, we showed in Section \ref{Subsection: higher derivative boost charge} that $K_\Upsilon=\frac{\hbar}{2\pi}S_{\text{dyn}}[\Upsilon]$. Therefore, the contribution to the entropy from entanglement across $\Upsilon$ can be written
\begin{equation}
	S_\Upsilon = \expval{S_{\text{dyn},\text{min}}}_A = \expval{\min_{\alpha \in G_\Upsilon}S_{\text{dyn}}[\alpha(\Upsilon)]}_A.
\end{equation}
In other words, this entropy is equal to the expectation value of the minimal dynamical entropy that the entangling surface can be deformed to have.

\section{Conclusions}
\label{Section: Discussion}

In this paper we considered density matrices in semiclassical gravity whose elements can be expressed as Euclidean path integrals, and asked what happens to the reduced density matrix associated to a subregion in the semiclassical limit. We found that the reduced density matrix elements are dominated by states for which a certain functional evaluated on the entangling surface at the boundary of the subregion is minimised with respect to deformations of that entangling surface. Moreover, we found that the von Neumann entropy of the subregion has a contribution associated to the entangling surface equal to $\frac{2\pi }{\hbar}$ multiplied by this functional.

In the case of general relativity the functional was equal to the area $\mathcal{A}[\Upsilon]$ of the entangling surface divided by $8\pi G$. The semiclassical entropy associated to the entangling surface was therefore equal to the expectation value of the minimal value of $\mathcal{A}[\Upsilon]$ divided by $4G\hbar$. This result is in agreement with the Ryu-Takayanagi conjecture.

For higher derivative theories of gravity, the functional was found to be proportional to a certain dynamical generalisation of the Wald entropy, $S_{\text{dyn}}[\Upsilon]$. The complete determination of the form of this dynamical entropy depends on a resolution and regularisation of the ordering ambiguities discussed in Section \ref{Section: Entangling surface charge}.

The next step is to find the correct regularisation procedure. Some clues in this direction may be found in the forms of certain generalisations of the Ryu-Takayanagi conjecture that exist in the literature~\citegeneralisedRT. In those generalisations, the correct entropy functional is given by the Wald entropy, plus some additional contributions involving higher order derivatives of the Lagrangian density, which are for example of the form $\pdv{L}{R_{abcd}}{R_{efgh}}$. Such contributions arise `anomalously' during the course of the generalised version of the Lewkowycz-Maldacena calculation. It should be noted that these anomalous contributions do not just modify the Wald entropy by the amount required for it to be equal to the Iyer-Wald dynamical entropy~\cite{Camps:2013zua,Jacobson:1993vj}. Anomalous terms in QFT are often a result of the need to properly regulate ordering ambiguities, which is exactly what we need to do here, so it is likely that agreement can be found by using similar methods to those papers.

Another possible future direction is the investigation of higher order quantum corrections to these results. Investigation in this direction is already ongoing~\cite{Faulkner:2013ana}, but the results developed in this paper may provide a new perspective. Consider general relativity in four dimensions, for which something particularly intriguing happens. The degree of freedom associated to deformations of the entangling surface will manifest itself in operator expectation values with a factor that looks like
\begin{equation}
	\int\Dd{\Upsilon} \exp(-\frac{\mathcal{A}[\Upsilon]}{4 G \hbar}).
\end{equation}
The integration is done over all possible deformations of $\Upsilon$. Suppose we want to view $\Upsilon$ as the worldsheet of a string. Then we recognise that this is just a path integral weighted by the Nambu-Goto action associated to that worldsheet, with string tension $\frac1{4G}$. In other words, the entangling surface does in fact behave quantum mechanically like a string in a curved background. It is very well-known that the classical string has a conformal symmetry on the worldsheet, and that this conformal symmetry only remains consistent after quantisation if the background metric obeys Einstein's equations~\cite{Callan:1985ia}. In this case, Einstein's equations are merely the equations of motion for the spacetime metric. Therefore, $\Upsilon$ can be expected to exhibit a 2D conformal symmetry that to a certain extent remains consistent quantum mechanically for free. 

Thus, even though our original assumptions made no reference to holography, we are forced to conjecture something like an ``entangling surface/CFT duality''! We should note that such a duality consists of a relation between a $D$-dimensional bulk and a $(D-2)$-dimensional boundary, which is one more codimension than usual. This may be related to results in~\cite{Jacobson:1995ab,Faulkner:2013ica}. It would be very interesting to understand whether this idea extends to higher dimensions and higher derivative theories of gravity.

Finally, we should mention the fact that our analysis has been done in a mostly Euclidean setting, whereas of course our reality is Lorentzian. We expect that our methods and conclusions should map directly to the Lorentzian case, but this has not been fully investigated. A complete understanding would require an analysis of the Wick rotation that is used to move between the Euclidean and Lorentzian path integrals. This would also reflect on the validity of the analytic continuation used to obtain \eqref{Equation: infinite boost boost charge}.

\section*{Acknowledgements}
\addcontentsline{toc}{section}{\protect\numberline{}Acknowledgements}
I am grateful to Malcolm Perry and Joan Camps for some useful discussions. I am also appreciative of the hospitality of the physics department at Harvard, where this work was carried out. This work was supported by a grant from STFC, and also grants from DAMTP and Clare College.

\printbibliography

\end{document}